\documentclass[twocolumn,pra,superscriptaddress,floatfix]{revtex4-1}

\usepackage{hyperref}
\usepackage{graphicx}
\usepackage{amsmath}
\usepackage{color}
\usepackage{units}
\usepackage[latin1]{inputenc}

\DeclareMathOperator{\CoV}{Cov} \newcommand{\VarOp}{\ensuremath{\Delta^2}}
\newcommand{\Var}[1]{\ensuremath{\VarOp {#1}}}

\newcommand{\VIBA}[1]{\ensuremath{\VarOp_{B|A} {#1}}}
\newcommand{\op}[1]{\ensuremath{\hat{#1}}}
\newcommand{\EPR}{\ensuremath{\mathcal{E}^2}}
\newcommand{\fig}[1]{Figure~\ref{fig:#1}}
\renewcommand{\P}{\ensuremath{P}}
\newcommand{\X}{\ensuremath{X}}

\begin{document}

\title{Strong Einstein-Podolsky-Rosen steering with unconditional entangled states}

\author{Sebastian Steinlechner}
\author{J{\"o}ran Bauchrowitz}
\author{Tobias Eberle}
\author{Roman Schnabel}
\affiliation{Institut f{\"u}r Gravitationsphysik, Leibniz Universit{\"a}t Hannover and Max-Planck-Institut f{\"u}r Gravitationsphysik (Albert-Einstein-Institut),  Callinstra{\ss}e 38, 30167 Hannover, Germany}

\begin{abstract}
In 1935 Schr\"odinger introduced the terms \emph{entanglement} and
\emph{steering} in the context of the famous gedanken experiment discussed by
Einstein, Podolsky, and Rosen (EPR).  Here, we report on a sixfold increase of
the observed EPR-steering effect with regard to previous experiments, as
quantified by the Reid-criterion. We achieved an unprecedented low conditional variance
product of about $0.04 < 1$, where 1 is the upper bound below which steering
is demonstrated.  The steering effect was observed on an unconditional
two-mode-squeezed entangled state that contained a total vacuum state
contribution of less than 8\%, including detection imperfections. Together with
the achieved high interference contrast between the entangled state and a
bright coherent laser field, our state is compatible with efficient
applications in high-power laser interferometers and fiber-based networks for
entanglement distribution.
\end{abstract}

\maketitle

Ever since the landmark article by A.~Einstein, N.~Podolsky, and B.~Rosen (EPR)
\cite{EPR1935} was published, entanglement has been demonstrated in a large
variety of physical systems \cite{AGR82,Ou1992,Hagley1997,Rowe2001,Blinov2004,Horo09}.
Recent theoretical and experimental work in quantum information
\cite{Wiseman07,Wiseman09,Wiseman2010} has sparked an interest in EPR
\emph{steering}. The term itself was already coined by E.
Schr\"{o}dinger in 1935 \cite{Sch35} in response to the original EPR
gedanken experiment.

The steering effect can be described as follows. One party, Alice, repeatedly
sends a defined physical system to another party, Bob.  She then proves to Bob
that she can predict his measurement outcomes on this system with more
precision than would be possible for any separable (classical) state.  In the
actual case of a shared \emph{steering} state, Bob would find rather broad
distributions of his measurement results for two non-commuting observables.
When he evaluates the discrepancies between Alice's predictions and his
measurement results, however, Bob realizes that the according mutual
uncertainties are less than allowed for a pure separable minimum uncertainty
state, as quantified by Heisenberg's uncertainty principle. It therefore seems
that Alice accomplishes a phase space \emph{steering} of the measurement
outcomes at Bob's side, merely by performing measurements on her subsystem. Of
course, other than the term \emph{steering} might suggest, in the course of
Alice's measurements no information is exchanged with Bob's subsystem. Only
when she sends her predictions, Bob realizes that her knowledge exceeds the
fundamental limit for a separable state and its non-commuting observables, such
as position and momentum.

EPR as well as Schr\"{o}dinger considered the example of an \emph{idealized},
pure bipartite entangled state of continuous variables (CV). In 1989,
M.\,D.\,Reid proposed a criterion for the EPR entanglement of
\emph{non-idealized} Gaussian, continuous variable systems, i.e.\ for
decohered, mixed Gaussian states \cite{Reid89}.  More recently,
E.\,G.\,Cavalcanti \emph{et al.}\ \cite{CRe07,CRDB11} derived criteria for
non-idealized \emph{discrete} systems, and H.\,M.\,Wiseman and co-workers
developed a general theory of experimental EPR-steering criteria applicable to
discrete as well as continuous-variable observables \cite{Wiseman07,Wiseman09}.
It was shown that the Reid criterion can be re-derived within this formalism.
The first demonstration of EPR steering was achieved by Z.\,Y.\,Ou \emph{et
al.}\ in 1992, followed by several other experiments in the CV regime
\cite{Zhang2000,Schori2002,Bowen2003,Laurat2004,Yonezawa2007,Keller2008,DAuria2009,
Wang2010,Eberle2011}. In 2003, W.\,P.\,Bowen \emph{et al.}\ experimentally
demonstrated in the Gaussian regime that EPR entanglement is indeed more
demanding than just establishing entanglement \cite{Bowen2003}.  In 2004,
J.\,C.\,Howell demonstrated the EPR paradox for continuous variables of single
photons using post-selection \cite{Howell2004}. A review on these experiments
is given in \cite{Reid2009}. EPR-steering was also observed in several
experiments with discrete variables, based on photon counting and
post-selection \cite{Wiseman2010,BESBCWP11,WRSLBWUZ11,SGABFWLCGNW11}.
Recently, we have shown that the intrinsic asymmetry in the steering
scenario can lead to one-way steering, where only Alice can steer Bob but not
vice-versa \cite{Haendchen2012}.

Several quantum information protocols for unconditional CV entanglement are
known. In each of these, quantitatively strong nonclassical properties are
required for useful implementations. For example, dense coding can increase the
capacity of a quantum-information channel, but requires more than \unit[4.8]dB
of two-mode squeezing to surpass classical schemes \cite{Braunstein2003}.
Analogously, the obtained secure bit rate in CV quantum key distribution
depends on the strength of the entanglement \cite{Furrer2012}.  In the field of
quantum metrology, recent theoretical and experimental works show promising
sensitivity improvements with unconditional entangled states
\cite{Polzik2010,Caves2012,Steinlechner2012,Genoni2013}. 

\begin{figure*}
\centering
\includegraphics[width=0.95\linewidth]{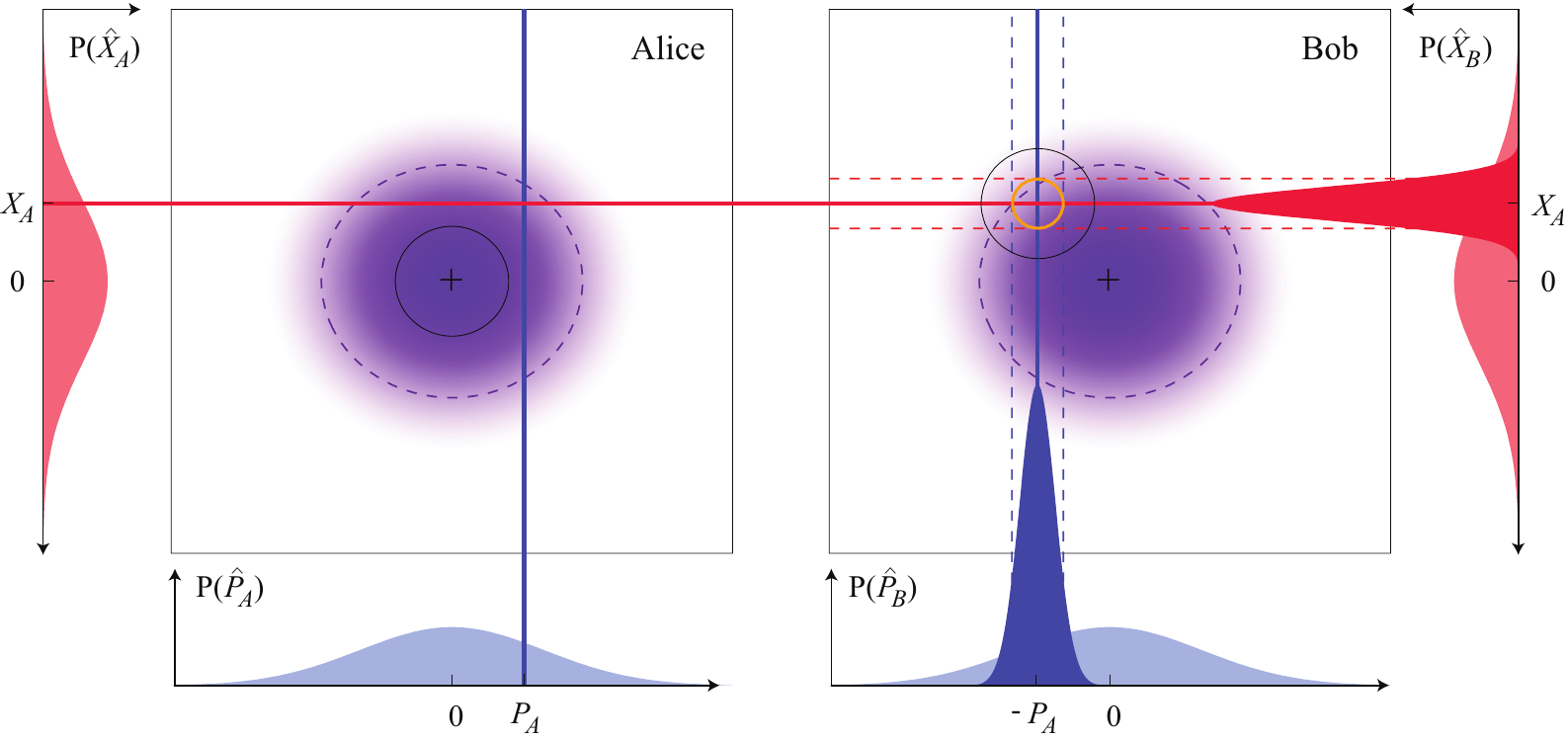}
\caption{(Color online) Graphical representation of the EPR-Reid criterion as a measure for
the continuous-variable EPR steering task. 
    Alice and Bob share an entangled state with probability distributions for
    each quadrature as given by the four broad projections for $\op \X$ and
    $\op \P$ in red and blue, respectively. From a measurement result $\X_A$
    (solid red line) at her side, Alice can predict the measurement outcome at
    Bob's detector with an uncertainty as given by the dashed red lines. The
    same applies to a measurement of $\P_A$, as indicated by the blue lines.
    The two narrow Gaussian projections and thus the area enclosed by the
    orange circle are measures for Alice's ability to predict Bob's measurement
    results.  For our experimental results, which are drawn here to scale, this conditional uncertainty circle has about
    one fifth the variance of the vacuum (black circle), which would be the
    lower uncertainty limit had Alice and Bob not shared an entangled state.
    }
\label{fig:epr-steering}
\end{figure*}

Here we report on the continuous observation of strong unconditional EPR
steering in the Gaussian regime using two-mode squeezed states. The EPR
steering strength is quantified to $\EPR = 0.041\pm0.005 < 1$, where unity is
the benchmark below which the steering effect is demonstrated \cite{Wiseman09}. This value is, to the
best of our knowledge, the strongest unconditional EPR entanglement measured to
date and a more than sixfold improvement over previous values \cite{Wang2010}.
The improvement was made possible by the recent advances in squeezed light
sources \cite{Eberle2010,Mehmet2011}, which allowed us to stably operate two
such sources simultaneously with detected squeezing values of about \unit[10]{dB}. A
special property of our EPR state is its low vacuum state contribution of just
8\%.  The EPR entanglement is contained in a well-defined $\text{TEM}_{00}$
mode and is therefore ideally suited for possible applications in optical
networks and high-power laser interferometers. 

In order to qualify and quantify the EPR-steering,
we use the EPR-Reid criterion \cite{Reid89} for the field quadratures $\op X$
and $\op P$,
\begin{equation}
\begin{split}
    \EPR_{B|A} &= \VIBA{\op \X} \cdot \VIBA{\op \P} \stackrel{!}< 1\,,
\end{split}
\label{eq:epr}
\end{equation}
where $\VIBA{\op O}= \min_{g_O} \Var{(\op O_B - g_O \op O_A)}$ is the
conditional variance and $\Var{\op O}$ denotes the variance of the observable
$\op O$.  The goal for Alice is to minimize the mutual uncertainty, i.e.\ the
conditional variances. If Alice is able to fulfill inequality~\eqref{eq:epr},
Bob will be convinced that they indeed share an entangled state. The critical
value of unity in \eqref{eq:epr} comes from the fact that classical
correlations between two beams can only be determined with at most the accuracy
of the vacuum's zero-point fluctuations. Throughout this paper, we normalize
the variance of this so-called vacuum noise to unity, $\Var{\op \X^\text{vac}}
= \Var{\op\P^\text{vac}} \equiv 1$.

Figure~(\ref{fig:epr-steering}) illustrates the EPR-Reid criterion used in
this work. It shows the joint two-mode quasi-probability distribution of the
EPR entangled state shared by Alice and Bob. The four broad Gaussian
distributions give the measurement statistics for measuring one or the other
non-commuting observable $\op \X$ and $\op \P$ on the respective subsystem $A$ or
$B$. The figure further illustrates the success of our experiment drawn to scale.
For both quadratures we achieved a conditional variance about five times
smaller than the minimal uncertainty product possible for separable states.
This is represented in the illustration by the orange circle on the right,
compared to the black circle.

\begin{figure}[tbp]
\centering
\includegraphics[width=0.75\linewidth]{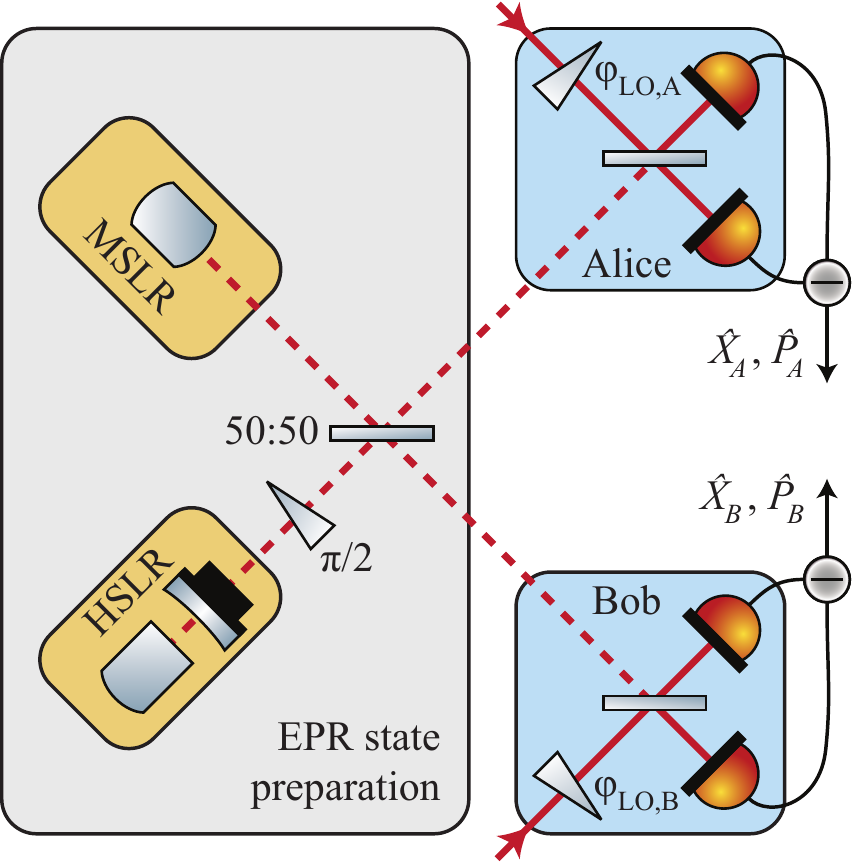}
\caption{(Color online) Schematic of the experimental setup. Two squeezed beams coming from
    a monolithic and a half-monolithic squeezed light resonator, {MSLR} and
   {HSLR}, are overlapped at a 50:50 beamsplitter, thereby
    producing a bi-partite EPR entangled state. The measurements at
   {Alice} and {Bob} are performed by balanced homodyne detection.
    }
\label{fig:setup}
\end{figure}

\emph{Experiment}---Figure \ref{fig:setup} shows a schematic of our
experimental setup. Two squeezed-light resonators (SLRs) produced
amplitude-quadrature squeezed light fields at a wavelength of \unit[1064]nm.
These were overlapped with a relative phase of $\pi/2$ on a 50:50 beam
splitter, thereby creating two-mode squeezing.  The quadrature amplitudes
$\op\X$ and $\op\P$ of the two output modes were detected with two balanced
homodyne detectors at Alice's and Bob's sites. Both detector outputs were passively
subtracted and then measured with a spectrum analyzer at a Fourier frequency of
\unit[5]MHz with a resolution bandwidth of \unit[300]kHz.

We used type\;I parametric down-conversion in periodically-poled potassium
titanyl phosphate (PPKTP) to generate the squeezed input fields. The two
squeezed light resonators differed in that one was monolithic ({MSLR}), with
cavity mirror coatings directly applied to the crystal's curved end faces,
while the other was half-monolithic (hemilithic) ({HSLR}), with the cavity
between one crystal surface and a separate, piezo-actuated mirror. In both
cases, one crystal surface was highly reflecting at both the \unit[1064]nm
fundamental wavelength and the \unit[532]nm pump field. The outcoupling mirrors
each had a power reflectivity of 90\% for the fundamental and 20\% for the
harmonic field.  Peltier elements were thermally connected to both nonlinear
crystals and used to temperature stabilize to the phase-matching condition.
Most of the main laser light coming from a Nd:YAG non-planar ring-oscillator
was converted into the pump field at \unit[532]nm for the parametric
down-conversion process in the SLRs. The second harmonic field was amplitude
stabilized with a Mach-Zehnder type interferometer and spatially filtered in a
mode-cleaning ring cavity. About \unit[60]mW pump power was needed for each
SLR.  Sub-milliwatt control fields carrying radio-frequency phase modulations
were injected into both squeezing resonators through their highly reflective
mirror coatings. They were used to lock the monolithic as well as the
hemilithic cavities on resonance by actuating the laser frequency and the
piezo-driven outcoupling mirror, respectively. The same phase modulations were
also employed to control the phase of the pump field and to lock the homodyne
detectors' quadrature angles.  A small fraction of light was tapped off the
main laser beam and spatially filtered. This beam was then divided to provide
about \unit[10]{mW} local oscillator power for each homodyne detector. The
homodyne detectors where equipped with custom-made photo diodes with a quantum
efficiency of $> 99\%$ and had a dark-noise clearance of about \unit[22]{dB}
below the vacuum noise.

\begin{figure}[tbp]
\centering
\includegraphics[width=0.95\linewidth]{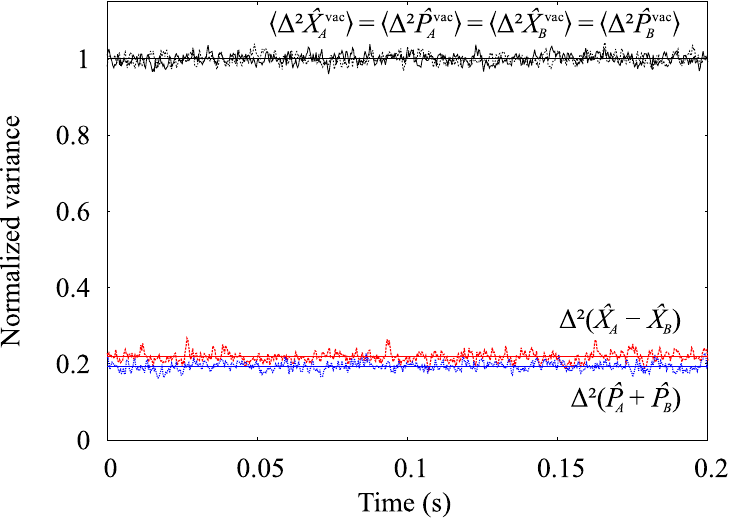}
\caption{(Color online) Variance of the sum and difference of the amplitude and phase
    quadratures at Alice and Bob. The traces were measured at a Fourier
    frequency of \unit[5]MHz and are normalized to the vacuum's zero-point
    fluctuations, $\Var\X^\text{vac} = \Var\P^\text{vac} \equiv 1$. Steering as
    characterized by the EPR-Reid criterion is clearly visible, here with
    scaling parameters $g_\X = 1$ and $g_\P = -1$, i.e.\ $\Var(\op X_A-\op
    X_B)\cdot\Var(\op P_A+\op P_B) < 1$. The traces were recorded with an RBW
    of \unit[300]{kHz}, and VBW of \unit[300]{Hz}.  Detection dark-noise was
    22\,dB below the vacuum noise and was not subtracted from the measurement
    data.}
\label{fig:noise-traces}
\end{figure}

\emph{Results}---A bipartite Gaussian entangled state is completely defined by
its covariance matrix $\gamma$. Given the entries of $\gamma$, the EPR-Reid
criterion \eqref{eq:epr} can be restated as \cite{Braunstein2003}
\begin{multline}
    \EPR_{B|A}= \biggl(\Var {\op X_B} - \frac{\CoV(\op X_A, \op X_B)^2}{\Var
    {\op X_A}}\biggr) \\
    \times \biggl(\Var {\op P_B} - \frac{\CoV(\op P_A, \op P_B)^2}{\Var {\op P_A}
    }\biggr) \,.
    \label{eq:epr2}
\end{multline}
We performed six different measurements on our entangled system in order to
partially reconstruct all relevant entries of its associated covariance matrix. These included the
amplitude and phase quadratures at Alice and Bob, respectively, and the
cross-correlations $\Var(\op X_A - \op X_B)$ and $\Var(\op P_A + \op P_B)$.
From the latter we calculated the covariances between measurements
at Alice and Bob via the identity
\begin{equation}
    \CoV(\op O_1, \op O_2) = \frac{1}{2}\Bigl(\Var(\op O_1+\op O_2) 
                             - \Var{\op O_1} - \Var{\op O_2}\Bigr)\,.
\end{equation}

An example of the measured traces is given in \fig{noise-traces}. These clearly
show the EPR steering effect,  $\EPR = 0.042 < 1$, where we chose the scaling
factors $g_\X = 1$ and $g_\P = -1$. The same traces also show the
inseparability of our system as expressed by the Duan criterion
\cite{Duan2000}, $\Var(\op X_A-\op X_B)+\Var(\op P_A+\op P_B) < 4$. From our
measurements the left side evaluates to $0.41$, thus falling below the Duan
threshold by almost a factor of ten.

The partially reconstructed covariance matrix reads
\begin{align}
    \gamma = \begin{pmatrix}
        18.41  &  (0)   & 18.09  &  (0)   \\
          (0)  & 35.49  &  (0)   & -34.95 \\
        18.09  &  (0)   & 17.98  &  (0)   \\
          (0)  & -34.95 &  (0)   & 34.61
    \end{pmatrix} \, . 
\end{align}
Each entry has an associated relative error of about 5\%. The $\op\P$
quadratures show roughly twice the variance of the $\op\X$ quadratures, which
can be attributed to almost \unit[3]dB more anti-squeezing produced in the
monolithic squeezed light resonator. The bracketed values where not measured
but instead `set' to zero by our experimental arrangement, because the
orientations of the squeezing ellipses as well as the local oscillator phases
of the BHDs were precisely controlled in such a way that no correlations were
introduced.  In principle these values can also be measured independently, as
we and others demonstrated before \cite{DAuria2005,DiGuglielmo2007,DAuria2009}. Such
measurements, however, are rather involved or even introduce additional optical
loss.  Note that if the bracketed zeros did not correspond to the actual
values, the generated entanglement strength would be underestimated but never
overestimated.

Inserting the entries of the covariance
matrix $\gamma$ into Eq.~\eqref{eq:epr2} yields $\EPR_{B|A} = 0.039$ for the
case where Alice steers the measurement outcome at Bob's detector.  This result
is slightly better than expected from \fig{noise-traces}, therefore we conclude
that initially, Alice's scaling parameters were not perfectly chosen.  The
reverse setup, $\EPR_{A|B}$, performs similar with an EPR value of $0.041$.

A comparison of $\gamma$ to a theoretical loss model yields an \emph{overall}
efficiency intrinsic to our physical system of $\xi = 92\%$.  Internal loss
of the squeezed light resonators (about $2.5\%$), propagation loss (1\%) and
imperfect mode overlap at the entanglement beam splitter (about 1.4\% loss due
to a fringe visibility of $0.993$)  lead to an EPR state preparation efficiency
of $\eta = (95\pm 1)\%$. We therefore obtain a homodyne detection efficiency of
$\xi/\eta = (97\pm 1)\%$, which is a reasonable value assuming a quantum
efficiency of our custom made photo-diodes of 99\%, additional propagation
loss of $0.6\%$, and again a fringe visibility of about $0.993$.

\emph{Discussion and conclusion}---In this work we demonstrated unconditional EPR-steering
using continuous measurements of position- and momentum-like
variables. We significantly improved the strength of the steering effect as
quantified by Eq.~\eqref{eq:epr} to $\EPR = 0.04$. This value corresponds to an
inseparability of $0.41 < 4$, according to Duan \emph{et al.}\ \cite{Duan2000}.

Our improvement is closely linked to the recent advances in squeezed light
sources, which we successfully transferred and applied to our entanglement
setup. The high quality of
our state's spatial mode makes it also applicable in more complex optical
networks, such as in interferometers, where an excellent mode-matching is
essential to achieve an overall high efficiency. From a fundamental point of view,
the setting used in this work is able to asymptotically approach the original
idealized \emph{gedanken experiment} considered by Einstein, Podolsky, Rosen
and Schr\"odinger.

We acknowledge support from the Centre for Quantum Engineering and Space-Time
Research (QUEST), the International Max Planck Research School on Gravitational
Wave Astronomy (IMPRS), the DFG Sonderforschungsbereich TR7, and the EU FP-7
project Q-ESSENCE, and helpful discussions with J\"org Duhme, Torsten Franz,
Melanie Meinders, and Reinhard Werner.

\end{document}